\documentclass[a4paper,nofootinbib,preprintnumbers,twocolumn,preprintnumbers,floatfix,superscriptaddress,prl,showpacs]{revtex4-1}

\usepackage{graphicx}
\usepackage{amssymb}
\usepackage{amsmath}
\usepackage{epstopdf}
\usepackage{graphics}
\usepackage[caption=false]{subfig}
\usepackage{float}
\usepackage{color}
\usepackage{hyperref}

\allowdisplaybreaks[1]

\newcommand{\ket}[1]{|#1\rangle}					

\newcommand{\abs}[1]{\left| #1 \right|} 
\newcommand{\avg}[1]{\langle #1 \rangle} 

\newcommand{\figref}[1]{Fig.~\ref{#1}}

\begin{document}

\title{Near Heisenberg limited atomic clocks in the presence of decoherence }

\author{J. Borregaard }
\author{A. S. S\o rensen}

\affiliation{QUANTOP, Niels Bohr Institute, University of Copenhagen, Blegdamsvej 17, DK-2100 Copenhagen �, Denmark}

\date{\today}

\begin{abstract}
The ultimate stability of atomic clocks is limited by the quantum noise of the atoms. To reduce this noise it has been suggested to use entangled atomic ensembles with reduced atomic noise. Potentially this can  push the stability all the way to the limit allowed by the Heisenberg uncertainty relation, which is denoted the Heisenberg limit. In practice, however, entangled states are often more prone to decoherence, which may prevent reaching this performance. Here we present an adaptive measurement protocol that in the presence of a realistic source of decoherence enables us to get near Heisenberg limited stability of atomic clocks using entangled atoms. The protocol may thus realize the full potential of entanglement for quantum metrology despite the detrimental influence of decoherence.  \end{abstract}


\maketitle

\begin{figure} 
\centering
\subfloat {\label{fig:atomicstate}\includegraphics[width=0.23\textwidth]{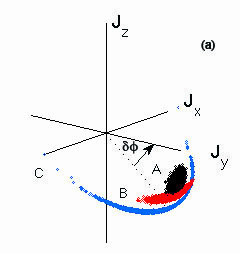}} 
\subfloat{\label{fig:ramseytime}\includegraphics[width=0.27\textwidth]{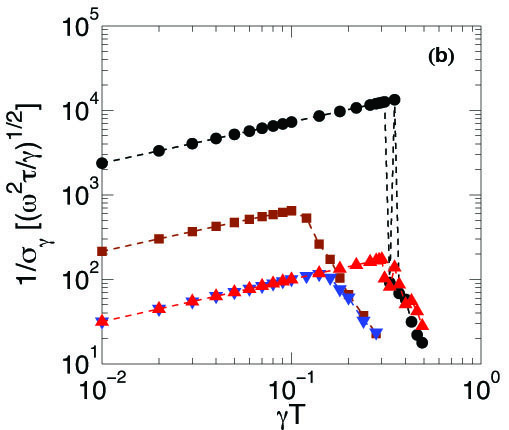}}
\caption{(Colour online) (a) The atomic state just before the measurement of $J_{z}$ for (A) uncorrelated atoms, (B) moderately squeezed atoms and (C) highly squeezed atoms. (b) Stability as a function of the Ramsey time ($\gamma T$) for $N=10^{5}$. $ \blacksquare$,$(\blacktriangledown)$ is the conventional protocol of Ref.~\cite{andre2004prl} for optimal squeezing (uncorrelated) atoms while $\bullet$,$(\blacktriangle)$ is the adaptive protocol for optimal squeezing (uncorrelated) atoms. The adaptive protocol allows for $\gamma T \sim 0.3$ while the conventional protocol only allows for $\gamma T \sim 0.1$ \cite{SM}.}
\end{figure}

Atomic clocks provide some of the most accurate time measurements in physics. One of the main limitations to the stability of atomic clocks is the quantum noise of the atoms, which leads to the standard quantum limit (SQL) where the stability scales as  $1/\sqrt{N}$ with $N$ being the number of atoms \cite{Santarelli1999,itano1993pra}. To overcome this noise it has been suggested to use entangled states with reduced atomic noise \cite{wineland1994pra,bollinger1996pra,andre2004prl,rosenband2012arxiv,lloyd}. Ultimately this may lead to a stability at the  Heisenberg limit where the resolution scales as $1/N$, and recently the first proof of principle experiments have  demonstrated these concepts experimentally \cite{leibfried2004science,Leroux,Appel,Anne,riedel,gross}.  In practice, however, entangled states are often more prone to decoherence, and to fully assess the advantage it is essential to study the  performance in the presence of decoherence \cite{huelgaprl1997}. In Ref.~\cite{andre2004prl} it was proven that entanglement can be used to improve the long-term stability of atomic clocks in the presence of the dominant practical source of decoherence, but the improvement identified was rather limited. Here we show that it is possible to obtain a large improvement in the stability of the clock by combining entanglement with an adaptive measurement protocol (inspired by  Ref.~\cite{wiseman2000prl, wisemannature2007}). With our adaptive measurement protocol the entangled states are not more sensitive to the decoherence than disentangled states (cf. \figref{fig:ramseytime}). As a consequence  the long term stability of the atomic clock can be improved almost to the Heisenberg limit even in the presence of decoherence. 

Many atomic clocks are operated by locking a local oscillator (LO) to an atomic transition via a feedback loop. The feedback is typically based on a measurement of the LO frequency offset $\delta\omega$ compared to the atomic transition through Ramsey spectroscopy~\cite{ramsey1956}. Here the atoms are first prepared in one of the two clock states by e.g. a laser pulse. During the Ramsey sequence the atoms interact with the LO field. This interaction consists of three parts; first the atoms are subject to a near-resonant $\pi/2$-pulse from the LO followed by the Ramsey time $T$ of free evolution, and finally another near-resonant $\pi/2$-pulse is applied to the atoms. During the free evolution the LO acquires a phase $\delta\phi=\delta\omega� T$ relative to the atoms. Due to the last $\pi/2$-pulses this phase can be measured as a population difference between the two clock levels. $\delta\omega$ can thus be estimated from the measurement and used for a feedback that steers the frequency of the LO to the atomic frequency. The stability of the clock will improve with $T$ since a longer $T$ improves the relative sensitivity of the frequency measurement. For current atomic fountain clocks, $T$ is limited by gravity and can hardly be varied~\cite{santarelliprl1999}. Here on the other hand we consider trapped particles, where $T$ can be increased until it  is limited by the decoherence in the system~\cite{chouprl2010,peik2006, nicholsonprl2012}. The long term stability thus depends on the nature of the decoherence. 

To take decoherence into account Ref.~\cite{huelgaprl1997} considered single atom dephasing. For this model Ref.~\cite{huelgaprl1997} showed that entanglement can not improve the stability of atomic clocks considerable (although an improvement is possible for non-Markovian noise~\cite{chin2012prl,yu2011}). A more realistic model of the decoherence was described in Ref.~\cite{andre2004prl} where the primary noise source is the frequency fluctuations of the LO~\cite{wineland1998jrnist}. In this work a small improvement in the long term stability, scaling as  $\sim N^{1/6}$, was identified for entangled atoms. Here we use the same decoherence model and disregard any decoherence of atoms, to show that entanglement and adaptive measurements may improve the performance and give near Heisenberg limited atomic clocks. Although the assumption of negligible atomic decoherence may be hard to fulfill for the highly entangled states considered here, our results highlight that there is no fundamental obstacle to reaching the Heisenberg limit. 
Another approach to increase the stability is to increase $T$~\cite{rosenbandarxiv2013,johannesarxiv,shiganjp2012}. In particular Ref. ~\cite{shiganjp2012} increases $T$ through a measurement protocol highly related to ours. However that work considers a scenario where the clock is limited by technical noise so that a direct comparison with our results is not possible. Which protocol is advantageous is thus an open question beyond the scope of this article.   

\begin{figure} 
\includegraphics[width=0.5\textwidth]{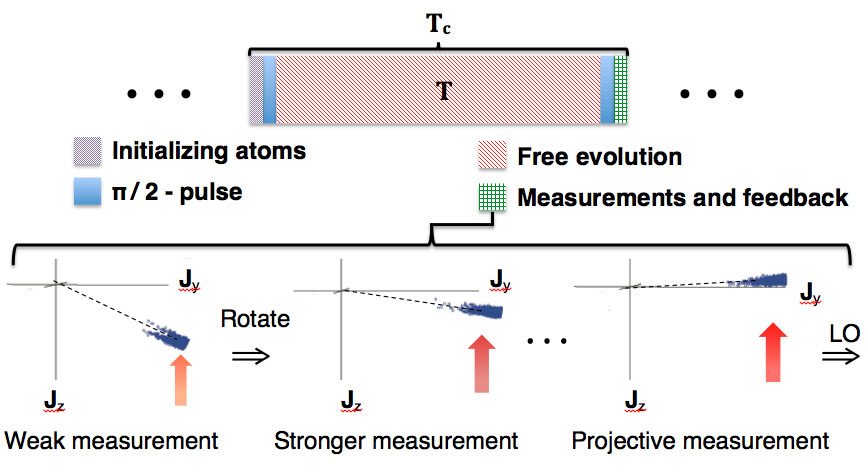} 
\caption{(Colour online) Operation of an atomic clock. A clock cycle of duration $T_{c}$ starts with initializing the atoms and ends with the measurements and feedback on the LO. The bottom part of the figure shows the adaptive protocol consisting of a series of weak measurements with intermediate feedback. The feedback seeks to rotate the atomic state to have mean spin almost along the $y$-axis before the final projective measurement and subsequent feedback on the LO.}
\label{fig:clockcycle}
\end{figure}

We consider an ensemble of $N$ two-level atoms, which we model as a collection of spin-1/2 particles with total angular momentum $\vec{J}$. The angular momentum operators $\hat{J}_{x,y,z}$ give the projections of $\vec{J}$ on the $x,y$ and $z$-axis.  The atoms are initially pumped to have a mean spin along the $z$-axis, $\avg{\hat{J}_{x}}=\avg{\hat{J}_{y}}=0$. After the Ramsey sequence the Heisenberg evolution of $\hat{J}_{x},\hat{J}_{y}$ and $\hat{J}_{z}$ is $\hat{J}_{1}(\delta\phi)=\hat{J}_{x},\hat{J}_{2}(\delta\phi)=\sin(\delta\phi)\hat{J}_{y}-\cos(\delta\phi)\hat{J}_{z}$ and $\hat{J}_{3}(\delta\phi)=\cos(\delta\phi)\hat{J}_{y}+\sin(\delta\phi)\hat{J}_{z}$. At the end of the Ramsey sequence $\hat{J}_{3}$ is measured and used to estimate $\delta\phi$. The $\hat{J}_{y}$ term in $\hat{J}_{3}$ results in the so called projection noise in the phase estimate $\sim \Delta J_{y}/{\abs{\avg{J_{z}}}}$. For uncorrelated atoms $\Delta\hat{J}_{y} \Delta\hat{J}_{x}=\avg{\hat{J}_{z}}/2\approx N/4$ and the projection noise causes the stability of the clock to scale as $\sim 1/\sqrt{N}$. For a spin squeezed state  \cite{kitagawa} the variance of $\hat{J}_{y}$ is reduced to obtain a better phase estimate. Such a spin squeezed state is depicted in \figref{fig:atomicstate}, which shows how the spin squeezed state looks like a "flat banana" on the Bloch sphere. 
The more we squeeze, the longer and more narrow the banana is and significant extra noise is added to the mean spin direction. For a phase estimate based on a direct measurement of $\hat{J}_{3}$ this gives an additional noise term  $\sim \delta\phi\Delta \hat{J}_{z}/{\abs{\avg{J_{z}}}}$. This extra noise limited the performance in Ref.~\cite{andre2004prl} if strongly squeezed states were used. We avoid this problem by using an adaptive scheme with weak measurements to make a rough estimate of $\delta\phi$ and then rotate the spins of the atoms such that the mean spin is almost along the $y$-axis. The flat banana depicted in \figref{fig:atomicstate} will then lie in the $xy$-plane and this will decrease the noise from $\Delta \hat{J}_{z}$ in subsequent measurements (see \figref{fig:clockcycle}). Having eliminated the noise from $\Delta\hat{J}_{z}$ we can allow strong squeezing in $\Delta\hat{J}_{y}$ and obtain near Heisenberg limited stability.

The operation of the clock consists of repeating the clock cycle illustrated in \figref{fig:clockcycle}. 
The total cycle duration $T_{c}$ will be larger than the period of free evolution due to the time spent on preparation and measurement of the atoms, and this dead time introduces Dick noise to the stability~\cite{dick}. To focus on the atomic noise we assume that the dead time is negligible ($T_{c} \sim T$) so that we can ignore the Dick noise. (Alternatively some clock based measurement can also be constructed which are immune to Dick noise~\cite{takamoto2011,hinkley,lodewyck2010}). This assumption is further discussed in the supplemental material \cite{SM}. We discretize time in the number of clock cycles ($k$) such that at time $t_{k}=kT$ the frequency correction $\Delta\omega(t_{k})=-\alpha \delta\phi^{e}(t_{k})/T$ is applied to the LO where $\alpha$ sets the strength of the feedback loop and $\delta\phi^{e}(t_{k})$ is the estimate of the accumulated phase $\delta\phi(t_{k})$ between time $t_{k-1}$ and $t_{k}$. The frequency offset of the LO at time $t_{k}$ is then $\delta\omega(t_{k})=\delta\omega_{0}(t_{k})+\sum_{i=1}^{k}\Delta\omega(t_{i})$, where $\delta\omega_{0}(t_{k})$ is the frequency fluctuation of the unlocked LO. The mean frequency offset after running for a period $\tau=lT$ ($l \gg 1$) is \cite{SM}
\begin{equation}
\delta\bar{\omega}(\tau)=\frac{1}{l}\sum_{k=1}^{l}\frac{\delta\phi(t_{k})-\delta\phi^{e}(t_{k})}{T},
\end{equation} 
resulting in the long term stability of the atomic clock: 
\begin{eqnarray}
\sigma_{\gamma}(\tau)&=&\avg{(\delta\bar{\omega}(\tau)/\omega)^{2}}^{1/2} \\
&=&\sqrt{\frac{1}{\tau\omega^{2}}}\left(\frac{1}{l}\frac{\avg{(\sum_{k=1}^{l}\delta\phi(t_{k})-\delta\phi^{e}(t_{k}))^{2}}}{T}\right)^{1/2} \label{eq:allan1/f}
\end{eqnarray}
We initially assume that the phase offset of the unlocked LO $\delta\phi_{0}$ is due to frequency fluctuations in the LO with a white noise spectrum. Later we will also consider the case where the fluctuations have a $1/f$ spectrum. For white noise we have $\avg{\delta\phi_{0}^{2}}=\gamma T$ ($\avg{\delta\phi_{0}}=0$) where $\gamma$ is a parameter characterizing the fluctuations. In the limit $\alpha\ll1$, the phases are uncorrelated \cite{SM} such that $\sigma_{\gamma}(\tau)=\sqrt{\gamma/\tau\omega^{2}}(\avg{(\delta\phi_{0}-\delta\phi^{e})^{2}}/\gamma T)^{1/2}$.  This expression shows that for fixed $\gamma$ and $\tau$ the stability of the clock only depends on how precisely we can estimate $\delta\phi_{0}$.  

We now describe our adaptive measurements in detail.  Our weak measurements is based on the strategy developed and demonstrated in Refs.~\cite{vuletic2010prl,Appel,Anne,polzikprl1998} where a light field dispersively interact with the spin and is subsequently measured. This is described by a Hamiltonian $H_{int}=-\chi_{1}\hat{J}_{3}\hat{X}_{1}$ where $\chi_{1}$ is the interactions strength and $\hat{X}_{1}$ is the canonical position operator of the light~\cite{hammerer2004pra,duan2000prl,kuzmich2000prl}. The measurement results in a rotation around $\hat{J}_{3}$ described by the rotation matrix $\mathbf{R}_{3}(\hat{\Pi}_{1})$, where $\hat{\Pi}_{1}=\Omega_{1}\hat{X}_{1}$. $\Omega_{1}=\chi_{1}\mu_{1}$ is the measurement strength and $\mu_{1}$ is the measurement time. The canonical momentum operators of the light  before $\hat{P}_{1}$ and after $\hat{P}_{1}^{'}$ the interaction are then related by $\hat{P}_{1}^{'}=\hat{P}_{1}-\Omega_{1}\hat{J}_{3}$. $\hat{P}_{1}^{'}$ is measured using homodyne detection~\cite{hammerer2010rmp} and the phase is estimated as $\delta\phi_{1}^{e}=\frac{-\beta_{1}\hat{P}_{1}^{'}}{\Omega_{1}\avg{\hat{J}_{z}}}$ where the factor $\beta_{1}$ is found from minimizing $\avg{(\delta\phi_{0}-\delta\phi_{1}^{e})^{2}}$. Based on the phase estimate we rotate the spin of the atoms around $\hat{J}_{1}$ in order to compensate for the extra noise added ($\Delta J_{z}$) by the spin squeezing. This is described by a rotation matrix $\mathbf{R}_{1}(\delta\phi_{1}^{e})$. The process can be iterated such that after $n\!-\!1$ weak measurements the Heisenberg evolution of the original operators ($\hat{J}_{1},\hat{J}_{2},\hat{J}_{3}$) is:
\begin{eqnarray} \label{eq:operatorupdate}
\left(\!\!\begin{array} {c}
\hat{J}_{1} \\
\hat{J}_{2} \\
\hat{J}_{3} \end{array}\!\! \right)_{n}\!\!\!\!=\mathbf{R}_{1}(\delta\phi_{n\!-\!1}^{e})\mathbf{R}_{3}(\hat{\Pi}_{n\!-\!1})...\mathbf{R}_{1}(\delta\phi_{1}^{e})\mathbf{R}_{3}(\hat{\Pi}_{1})\!\!\left(\!\!\begin{array} {c}
\hat{J}_{1} \\
\hat{J}_{2} \\
\hat{J}_{3} \end{array} \!\!\right) \quad
\end{eqnarray}
The final measurement is assumed to be a projective measurement and the final phase estimate $\delta\phi_{n}^{e}$ is thus $\delta\phi_{n}^{e}=\frac{\beta_{n}\hat{J}_{3,n}}{\avg{\hat{J}_{z}}}$. The factors of $\beta_{i}$ in the phase estimates are found be minimizing $\avg{(\delta\phi_{0}-\sum_{j=1}^{i}\delta\phi_{j}^{e})^{2}}$ with respect to $\beta_{i}$ after each measurement.  The final estimate of $\delta\phi_{0}$ at the end of the measurement sequence is $\delta\phi^{e}=\sum_{i=1}^{n}\delta\phi_{i}^{e}$ where $\delta\phi_{i}^{e}$ is the phase estimate after the $i$'th measurement. 

We will now show semi-analytically that the measurement strategy in equation \eqref{eq:operatorupdate} allows for near Heisenberg limited stability. For simplicity we set all $\beta_{i}=1$ in our analytical calculations. After $j$ weak measurements the difference between the true phase and the estimated phase $\delta\Phi_{j}$ is:
\begin{equation}
\delta\Phi_{j}=\delta\phi_{0}-\sum_{i=1}^{j}\delta\phi_{i}^{e}=\delta\phi_{0}-\sum_{i=1}^{j-1}\delta\phi_{i}^{e}-\delta\phi_{j}^{e}.
\end{equation}
Using equation \eqref{eq:operatorupdate} to get an expression for $\delta\phi_{j}^{e}$ and the fact that $\delta\Phi_{j\!-\!1}=\delta\phi_{0}-\sum_{i=1}^{j-1}\delta\phi_{i}^{e}$, we can express the phase error as
\begin{eqnarray} 
\delta\Phi_{j}\!\!&\approx&\!\!\delta\Phi_{j\!-\!1}(1\!-\!\hat{J}_{z}/\avg{\hat{J}_{z}})\!-\!(\hat{J}_{y}+\delta\hat{J}_{3,j}\!-\!\hat{P}_{j}/\Omega_{j})/\avg{\hat{J}_{z}}, \label{eq:offset}
\end{eqnarray}
where we have assumed $\delta\Phi_{j\!-\!1}\!\!\ll \!\!1$. The first term in \eqref{eq:offset} gives a contribution $\sim \delta\Phi_{j\!-\!1}\Delta J_{z}/\avg{\hat{J}_{z}} $ to $\sigma_{\gamma}(\tau)$ from the noise in the mean spin direction as discussed previously. Note that this term is proportional to the phase estimation error at the previous measurement stage, since it depends on how well the 'banana' in \figref{fig:atomicstate} is rotated into the $xy$-plane. For a useful adaptive protocol $\delta\Phi_{j\!-\!1}$ gets smaller for growing $j$ and the noise that enters through $\Delta\hat{J}_{z}$ is reduced. The last terms in Eq.\eqref{eq:offset} gives the noise from $\Delta \hat{J}_{y}$, the accumulated back action of the previous measurements ($\avg{\delta J_{3,j}^{2}}$), and the noise from the incoming light in the measurement ($\Delta\hat{P}_{j}^{2}=\avg{\hat{P}_{j}^{2}}$).
 
The stronger a measurement is, the less noise is added through $\Delta \hat{P}_{j}^{2}/\Omega^{2}_{j}$ since the measurement is more precise. Any imprecision $\Delta \hat{P}_{i<j}^{2}/\Omega^{2}_{i<j}$ from previous measurements is contained in $\delta\Phi_{j\!-\!1}$ and is corrected for in the subsequent stages of the protocol, which estimate how well we corrected the phase in previous measurements. This means that we can initially work with weak measurements, which only give a rough estimate since later stronger measurements correct for the imprecision in the initial measurements. 

The accumulated back action noise $\delta \hat{J}_{3,j}$ originates from the disturbance caused by the measurements. The measurements add noise in $\hat{J}_{1},\hat{J}_{2}$, which is mixed into $\hat{J}_{3}$ when the atomic state is rotated to have mean spin almost along the $y$-axis. From equation \eqref{eq:operatorupdate} the dominant term in $\delta \hat{J}_{3,j}$ is found to be $\delta \hat{J}_{3,j}\sim \sum_{i=1}^{j-1}\delta\phi_{i}^{e}\Omega_{i}\hat{X}_{i}\hat{J}_{x}$ \cite{SM}. The stronger a measurement is, the more noise is added to the stability. For a useful adaptive protocol however, $\delta\phi^{e}_{i}$ gets smaller for growing $i$, which means that the $i$'th measurement can be stronger than the previous $(i\!-\!1)$'th measurements without adding more noise to the stability. 

Above we have argued that we can suppress the noise terms originating from $\Delta J_{z}$, $\Delta \hat{P}_{j}$, and $\delta \hat{J}_{3,j}$ using an adaptive protocol with weak initial measurements. A remaining question is how well this suppression work. This is considered in detail in the supplemental material \cite{SM}. To be specific, we consider spin squeezed states of the form $\ket{\psi(\kappa)}=\mathcal{N}(\kappa)\sum_{m}(-1)^{m}e^{-(m/\kappa)^{2}}\ket{m}$, where $\ket{m}$ are eigenstates of $\hat{J}_{y}$ with eigenvalue $m$, $\mathcal{N}(\kappa)$ is a normalization constant and the sum is from $-J$ to $J$ where $J=N/2$ is the total angular momentum quantum number. This form gives a simple family of states characterized by a single parameter ($\kappa$), which can extrapolate between uncorrelated states $\kappa=\sqrt{N}$ and highly squeezed states approaching the $|m=0\rangle$ Fock state $\kappa\rightarrow 0$. It may be possible to identify more optimal states~\cite{andersprl2000} but this simple form is sufficient for our present purpose. We furthermore assume that the probe light has vacuum statistics. As an upper limit of the stability we find that for $n\gtrsim 3\log(N)$ weak measurements, using a spin squeezed state with $\kappa\sim \log\sqrt{N}+2$ and choosing a measurement strategy with $\Omega_{i} \sim N^{-1+i/(n+1)}$  we can suppress other noise terms so that the measurement is eventually limited by $\Delta \hat{J}_{y}$. $\sigma_{\gamma}(\tau)$ will then be $\sim (2/N+\log\sqrt{N}/N)/\sqrt{\gamma T}$ (in units of $\sqrt{\gamma/\tau\omega^{2}}$) for $N\gg1$. It is seen that for $N=10^{6}$ the upper limit of the stability will differ from the Heisenberg limit, $\sigma_{\gamma}(\tau)=(1/N)/\sqrt{\gamma T}$ by a factor of $\sim5$. 

The above upper limit shows that we can be near the Heisenberg limit, but to get the optimal stability we numerically minimized $\sigma_{\gamma}$. We simulated an atomic clock with a LO subject to white Gaussian noise for atom numbers in the range $N=100$ to $N=10^{6}$ \cite{SM}. For $N\leq1000$ we simulated the full quantum evolution (we denote this as 'full quantum simulation') and for $N>1000$ we approximated the probability distributions of $\hat{J}_{x,y,z}$ with Gaussian distributions with moments calculated from $\ket{\psi(\kappa)}$ in the approximation $N\gg1$ (we denote this as 'Gaussian simulation'). 

An example of the results are shown in  \figref{fig:ramseytime}, which demonstrates that there is a significant improvement by using a spin squeezed state compared to uncorrelated atoms. With the adaptive protocol, the Ramsey time can be as large for highly entangled states as for disentangled states and there is thus no difference in the relevant coherence time.  Furthermore because the adaptive protocol can determine phases $\lesssim \pi$ it allows longer interrogation times $\gamma T\lesssim 0.3$ than the conventional protocol $\gamma T\lesssim 0.1$, which begins to give ambiguous results for phases $\sim \pi/2$.  
 \begin{figure} 
\centering
\subfloat {\label{fig:white}\includegraphics[width=0.25\textwidth]{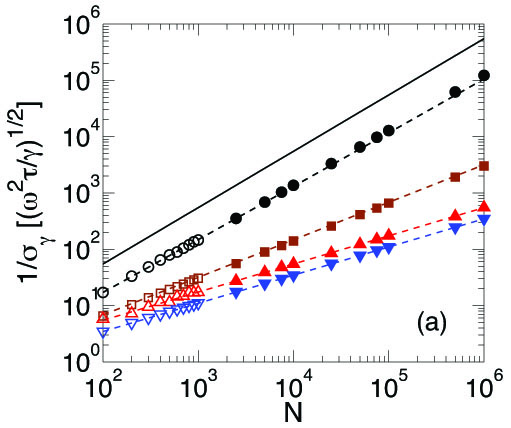}} 
\subfloat{\label{fig:pink}\includegraphics[width=0.25\textwidth]{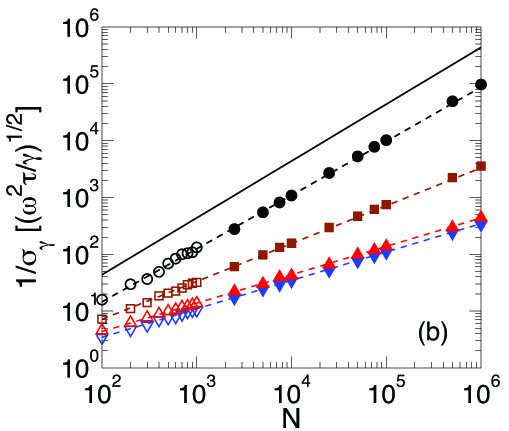}}
\caption{(Colour online) Optimized stability of an atomic clock for a LO subject to (a) white noise and (b) $1/f$-noise. $\circ, \square,\vartriangle,\triangledown$ are the full quantum simulation while $\bullet,\blacksquare,\blacktriangle, \blacktriangledown$ are the Gaussian simulation. The Gaussian simulation can be extended down to $N=100$, which give more or less identical results to the full quantum simulation. $\circ$,$\bullet$ ($\blacktriangle$,$\vartriangle$) are the adaptive scheme 
and $\square$, $\blacksquare$ ($\triangledown$,$\blacktriangledown$) are the conventional protocol with (without) entanglement. The dotted lines are the analytical results and the solid line is the Heisenberg limit for the maximal Ramsey time $\gamma T=0.3$ (a) and $\gamma T=0.2$ (b).}
\label{fig:optim}
\end{figure}

We have numerically minimized $\sigma_{\gamma}(\tau)$ in the degree of squeezing, the number of weak measurements, the Ramsey time, and the strengths of the measurements. \figref{fig:optim}(a) shows the result of the optimization for both the adaptive protocol and the conventional protocol with/without squeezing.  The adaptive protocol gives a significant improvement compared to using uncorrelated atoms resulting in near Heisenberg limited stability. The numerical calculations also agree nicely with the analytical calculations \cite{SM}. As noted above the adaptive protocol allows for a longer Ramsey time than the conventional protocol, which gives an improvement of roughly a factor 1.6 for uncorrelated atoms. 

 So far we have assumed white noise in the LO. In practice the noise of the LO is however more likely to have a nontrivial spectrum like $1/f$-noise. We have therefore repeated the numerical optimization with $1/f$-noise in the LO for which $\avg{\omega(f)\omega(f')}=\delta({f+f'})\gamma^{2}/f$ and the results are shown in \figref{fig:optim}(b) \cite{SM}. The improvement obtained using the adaptive scheme with correlated atoms persists also for $1/f$ noise. Again near Heisenberg limited stability is obtained using the adaptive protocol. The longer Ramsey time of the adaptive scheme compared to projective measurements gives an improvement of roughly a factor 1.3 for uncorrelated atoms.  
 
In conclusion we have developed an adaptive measurement protocol which allows operating atomic clocks near the Heisenberg limit using entangled spin squeezed ensembles of atoms. These results clearly demonstrate that entanglement can be an important resource for quantum metrology. Importantly our results are obtained under realistic assumptions where we account for the dominant source of noise in practice.  We find that in this situation we can gain nearly the full potential of entanglement estimated without accounting for decoherence.  Furthermore the adaptive protocol allows for a higher Ramsey time, which gives an improvement even for uncorrelated atoms.    

We gratefully acknowledge the support of the Lundbeck Foundation and the Danish National Research Foundation through QUANTOP. The research leading to these results has received funding from the European Research Council under the European Union's Seventh Framework Programme (FP/2007-2013) / ERC Grant Agreement n. 306576.

\newpage

\onecolumngrid

\renewcommand{\theequation}{S\arabic{equation}}
\renewcommand{\thefigure}{S\arabic{figure}}

\section{Supplemental material: Near Heisenberg limited atomic clocks in the presence of decoherence}

This supplemental material to our article "Near Heisenberg limited atomic clocks in the presence of decoherence" consists of two parts. In the first part we will go through the adaptive protocol in detail and describe the various noise terms that affect the stability of the clock. We show semi-analytically that the adaptive protocol gives near Heisenberg limited clocks with a stability scaling at most like $\log{N}/N$. The second part describes the details of our numerical simulations of an atomic clock. We describe the  subtleties of simulating an atomic clock running for a long but finite time and show how the feedback on the LO effectively locks it to the atoms.    

\section{Scaling of the stability}

In the article we describe how the adaptive protocol consists of a series of weak measurements of the atomic spin with intermediate feedback on the atoms. After a weak measurement we estimate the phase of the LO relative to the atoms as $\delta\phi_{i}^{e}=\frac{-\beta_{i}\hat{P}_{i}^{'}}{\Omega_{i}\avg{\hat{J}_{z}}}$ where $\hat{P}_{i}^{'}$ is the canonical momentum operator of the light after the interaction with the atoms and $\Omega_{i}$ is the measurement strength. To simplify the equations we set $\beta_{i}=1$ throughout this section. Note that this choice is not ideal and the true performance of the clock will thus be better than what we estimate here. Later we will argue that  for white noise and in the limit of a weak feedback ($\alpha\ll 1$) we can determine the stability of the LO from looking at the the error between the estimated phase and the true phase of the LO for each Ramsey sequence independently. We thus consider a Ramsey sequence where the LO acquires a phase $\delta\phi_{0}$ relative to the atoms during the free evolution. $\delta\phi_{0}$ is subsequently estimated through $n-1$ weak measurements with intermediate feedback on the atoms and a final projective measurement.  We define the operators
\begin{equation}
\hat{J}_{1}=\hat{J}_{x}, \qquad \hat{J}_{2}(\theta)=\sin(\theta)\hat{J}_{y}-\cos(\theta)\hat{J}_{z}, \qquad  \hat{J}_{3}(\theta)=\sin(\theta)\hat{J}_{z}+\cos(\theta)\hat{J}_{y}
\end{equation} 
such that the outcome of the Heisenberg evolution of $\hat{J}_{x},\hat{J}_{y}$ and $\hat{J}_{z}$ during the Ramsey sequence is $\hat{J}_{1},\hat{J}_{2}(\delta\phi_{0})$ and $\hat{J}_{3}(\delta\phi_{0})$. In the Heisenberg picture the first weak measurement results in a rotation of the atomic spin described by a rotation matrix $\mathbf{R}_{3}(\hat{\Pi}_{1})$, where $\hat{\Pi}_{1}=\Omega_{1}\hat{X}_{1}$ as described in the article ($\hat{X}_{1}$ is the canonical position operator of the probe light). The subsequent feedback on the atoms is a rotation described by the rotation matrix $\mathbf{R}_{1}(\delta\phi_{1}^{e})$ where $\delta\phi_{1}^{e}$ is the phase estimate based on the measurement result. Thus the Heisenberg evolution of the operators after the first weak measurement and subsequent feedback is: 
 \begin{eqnarray} \label{eq:operatorupdate}
\left(\!\!\begin{array} {c}
\hat{J}_{1} \\
\hat{J}_{2} \\
\hat{J}_{3} \end{array}\!\! \right)_{2}\!\!\!\!&=&\mathbf{R}_{1}(\delta\phi_{1}^{e})\mathbf{R}_{3}(\hat{\Pi}_{1})\!\!\left(\!\!\begin{array} {c}
\hat{J}_{1} \\
\hat{J}_{2}(\delta\phi_{0}) \\
\hat{J}_{3}(\delta\phi_{0}) \end{array} \!\!\right) \\
&=&\left(\!\!\begin{array} {ccc}
\cos\hat{\Pi}_{1} & -\sin\hat{\Pi}_{1} & 0 \\
\cos\delta\phi_{1}^{e}\sin(\hat{\Pi}_{1}) & \cos(\delta\phi_{1}^{e})\cos(\hat{\Pi}_{1}) &-\sin(\delta\phi_{1}^{e}) \\
\sin(\delta\phi_{1}^{e})\sin(\hat{\Pi}_{1})  & \sin(\delta\phi_{1}^{e})\cos(\hat{\Pi}_{1}) & \cos(\delta\phi_{1}^{e}) \end{array}\!\! \right)\!\!\left(\!\!\begin{array} {c}
\hat{J}_{1} \\
\hat{J}_{2}(\delta\phi_{0}) \\
\hat{J}_{3}(\delta\phi_{0}) \end{array} \!\!\right) \\
&=&\!\!\left(\!\!\begin{array} {c}
\hat{J}_{1} \\
\hat{J}_{2}(\delta\phi_{0}-\delta\phi_{1}^{e}) \\
\hat{J}_{3}(\delta\phi_{0}-\delta\phi_{1}^{e})\end{array} \!\!\right)+\!\!\left(\!\!\begin{array} {c}
\delta\hat{J}_{1,2} \\
\delta\hat{J}_{2,2}  \\
\delta\hat{J}_{3,2} \end{array} \!\!\right)
\end{eqnarray} 
where the operators
\begin{eqnarray}
\delta\hat{J}_{1,2}&=&(\cos(\hat{\Pi}_{1})\!-\!1)\hat{J}_{1}\!-\!\sin(\hat{\Pi}_{1}) \hat{J}_{2}(\delta\phi_{0})\\
\delta\hat{J}_{2,2}&=&\cos(\delta\phi_{1}^{e})\sin(\hat{\Pi}_{1})\hat{J}_{1}\!+\cos(\delta\phi_{1}^{e})(\!\cos(\hat{\Pi}_{1})\!-\!1) \hat{J}_{2}(\delta\phi_{0}) \\
\delta\hat{J}_{3,2}&=&\sin(\delta\phi_{1}^{e})\sin(\hat{\Pi}_{1})\hat{J}_{1}\!+\sin(\delta\phi_{1}^{e})(\!\cos(\hat{\Pi}_{1})\!-\!1) \hat{J}_{2}(\delta\phi_{0})  
\end{eqnarray}
describe the noise due to the back action of the measurement. Note that the back action also affects $\hat{J}_{3}$ even though $\hat{J}_{3}$ is conserved and hence unaffected during the measurement. This is because the  rotation during the feedback mixes back action noise into $\hat{J}_{3}$.   

The process is now iterated such that the Heisenberg evolution after $n-1$ weak measurements and subsequent rotations is
\begin{eqnarray} \label{eq:Hevolution}
\left(\!\!\begin{array} {c}
\hat{J}_{1} \\
\hat{J}_{2} \\
\hat{J}_{3} \end{array}\!\! \right)_{n}\!\!\!\!&=&\!\!\left(\!\!\begin{array} {c}
\hat{J}_{1} \\
\hat{J}_{2}(\delta\phi_{0}-\sum_{i=1}^{n\!-\!1}\delta\phi_{i}^{e}) \\
\hat{J}_{3}(\delta\phi_{0}-\sum_{i=1}^{n\!-\!1}\delta\phi_{i}^{e})\end{array} \!\!\right)+\!\!\left(\!\!\begin{array} {c}
\delta\hat{J}_{1,n} \\
\delta\hat{J}_{2,n}  \\
\delta\hat{J}_{3,n} \end{array} \!\!\right)
\end{eqnarray} 
where the iterative expressions for $\delta\hat{J}_{1,j-1},\delta\hat{J}_{2,j-1}$ and $\delta\hat{J}_{3,j-1}$ are
\begin{eqnarray}
\delta\hat{J}_{1,j}&=&(\cos(\hat{\Pi}_{j\!-\!1})\!-\!1)\hat{J}_{1}\!-\!\sin(\hat{\Pi}_{j\!-\!1}) \hat{J}_{2}{\left( \delta\phi_{0}\!-\!\sum_{i=1}^{j\!-\!1}\delta\phi_{i}^{e}\right)}+\cos(\hat{\Pi}_{j\!-\!1})\delta\hat{J}_{1,j-1}\!-\!\sin(\hat{\Pi}_{j\!-\!1})\delta\hat{J}_{2,j-1}  \label{eq:noiseoperators1} \\
\delta\hat{J}_{2,j}&=&\cos(\delta\phi_{j\!-\!1}^{e})\sin(\hat{\Pi}_{j\!-\!1})\hat{J}_{1}\!+\cos(\delta\phi_{j\!-\!1}^{e})(\!\cos(\hat{\Pi}_{j\!-\!1})\!-\!1) \hat{J}_{2}{\left(\delta\phi_{0}\!-\!\sum_{i=1}^{j\!-\!1}\delta\phi_{i}^{e}\right)}+\cos(\delta\phi_{j\!-\!1}^{e})\sin(\hat{\Pi}_{j\!-\!1})\delta\hat{J}_{1,j-1}\\
&+&\cos(\delta\phi_{j\!-\!1}^{e})\cos(\hat{\Pi}_{j\!-\!1})\delta\hat{J}_{2,j-1}\!-\!\sin(\delta\phi_{j\!-\!1}^{e})\delta\hat{J}_{3,j-1} \nonumber \\
\delta\hat{J}_{3,j}&=&\sin(\delta\phi_{j\!-\!1}^{e})\sin(\hat{\Pi}_{j\!-\!1})\hat{J}_{1}\!+\sin(\delta\phi_{j\!-\!1}^{e})(\!\cos(\hat{\Pi}_{j\!-\!1})\!-\!1) \hat{J}_{2}{\left(\delta\phi_{0}\!-\!\sum_{i=1}^{j\!-\!1}\delta\phi_{i}^{e}\right)} +\sin(\delta\phi_{j\!-\!1}^{e})\sin(\hat{\Pi}_{j\!-\!1})\delta\hat{J}_{1,j-1} \nonumber \\
&+&\sin(\delta\phi_{j\!-\!1}^{e})\cos(\hat{\Pi}_{j\!-\!1})\delta\hat{J}_{2,j-1}\!+\!\cos(\delta\phi_{j\!-\!1}^{e})\delta\hat{J}_{3,j-1}   \label{eq:noiseoperators2}
\end{eqnarray}
with $\delta\hat{J}_{1,1}=\delta\hat{J}_{2,1}=\delta\hat{J}_{3,1}=0$. 
In the final \emph{projective} measurement we measure $\hat{J}_{3,n}$ and obtain a phase estimate $\delta\phi_{n}^{e}=\hat{J}_{3,n}/\avg{\hat{J}_{z}}$. Our final estimate of $\delta\phi_{0}$ is then $\delta\phi^{e}=\sum_{i=1}^{n}\delta\phi_{i}^{e}$, i.e. $\delta\phi^{e}_{i}$ refers to a phase estimate during the adaptive measurement sequence while $\delta\phi^{e}$ is the final phase estimate at the end of the adaptive measurement sequence. The difference between $\delta\phi^{e}$ and the true phase is
\begin{equation}
\delta\Phi_{n}=\delta\phi_{0}-\sum_{i=1}^{n}\delta\phi_{i}^{e}=\delta\phi_{0}-\sum_{i=1}^{n-1}\delta\phi_{i}^{e}-\delta\phi_{n}^{e}.
\end{equation}
Using equation \eqref{eq:Hevolution} we can express this using the previous phase estimation error ($\delta\Phi_{n\!-\!1}$) and the last measurement
\begin{eqnarray} \label{eq:estimate}
\delta\Phi_{n}&=&\delta\Phi_{n\!-\!1}-\left(\sin(\delta\Phi_{n\!-\!1})\hat{J}_{z}+\cos(\delta\Phi_{n\!-\!1})\hat{J}_{y}+\delta\hat{J}_{3,n}\right)/\avg{\hat{J}_{z}} \label{eq:estimate1} \\
&\approx&\!\!\delta\Phi_{n\!-\!1}(1\!-\!\hat{J}_{z}/\avg{\hat{J}_{z}})\!-\!\hat{J}_{y}/\avg{\hat{J}_{z}}-\delta\hat{J}_{3,n}/\avg{\hat{J}_{z}} \label{eq:estimate2}
\end{eqnarray}
where $\delta\Phi_{n\!-\!1}=\delta\phi_{0}-\sum_{i=1}^{n-1}\delta\phi_{i}^{e}$ and we have assumed  $\delta\Phi_{n\!-\!1}\ll$1 to expand the sine and cosine. As noted above we will argue later that the stability of the LO will be given by the phase error $\avg{\delta\Phi_{n}^{2}}^{1/2}$ for each Ramsey sequence independently. We shall therefore now explore the limitations to the stability from the noise terms in $\avg{\delta\Phi_{n}^{2}}^{1/2}$ in this limit.

Equation \eqref{eq:estimate2} contains three kinds of noises that limit the stability. The first term is due to the uncertainty in $\hat{J}_{z}$ ($\Delta J_{z}$) and is proportional to $\delta\Phi_{n\!-\!1}$. This is the noise that limited the performance of the conventional protocol for strongly squeezed states in Ref. \cite{andre2004prl} and the idea of the adaptive protocol is to eliminate this noise by minimizing $\delta\Phi_{n\!-\!1}$ through many measurements and feedback. The second term is the noise in $\hat{J}_{y}$, which is the term we want to  decrease by squeezing the atomic spin. The final term is the accumulated back action of the measurements. The weaker the measurements are the smaller the contribution from this term will be.  

We now analyze the various terms in $\avg{\delta\Phi_{n}^{2}}$ in more detail. The estimated phase in the $j$'th weak measurement is
\begin{equation} \label{eq:estimatei}
 \delta\phi_{j}^{e}=\left(\sin(\delta\Phi_{j\!-\!1})\hat{J}_{z}+\cos(\delta\Phi_{j\!-\!1})\hat{J}_{y}+\delta\hat{J}_{3,j}-\hat{P}_{j}/\Omega_{j}\right)/\avg{\hat{J}_{z}}.
\end{equation}
Using equation \eqref{eq:noiseoperators1}-\eqref{eq:noiseoperators2} and equation \eqref{eq:estimate2}-\eqref{eq:estimatei} we can express the dominant contributions to the stability due to the noise in $\hat{J}_{z}$ as 
\begin{eqnarray}
\avg{\delta\phi_{0}^{2}}&&\avg{(1\!-\!\hat{J}_{z}/\avg{\hat{J}_{z}})^{2n}} \label{eq:thetanoise1} \\
+2\avg{\delta\phi_{0}(\sin(\delta\phi_{0})-\delta\phi_{0})}&&\avg{(1\!-\!\hat{J}_{z}/\avg{\hat{J}_{z}})^{2n-1}\hat{J}_{z}/\avg{\hat{J}_{z}}} \\
+\avg{(\sin(\delta\phi_{0})-\delta\phi_{0})^{2}}&&\avg{(1\!-\!\hat{J}_{z}/\avg{\hat{J}_{z}})^{2n-2}\hat{J}_{z}^{2}/\avg{\hat{J}_{z}}^{2}}. \label{eq:thetanoise3}
\end{eqnarray}
Here "dominant" refers to decreasing slowest with $N$. 

The above expressions are independent of the atomic state but we will now focus on a specific type of states in order to treat the system in more detail. We consider spin squeezed states of the form $\ket{\psi(\kappa)}=\mathcal{N}(\kappa)\sum_{m}(-1)^{m}e^{-(m/\kappa)^{2}}\ket{m}$, where $\ket{m}$ are eigenstates of $\hat{J}_{y}$ with eigenvalue $m$, $\mathcal{N}(\kappa)$ is a normalization constant and the sum is from $-J$ to $J$ where $J=N/2$ is the total angular momentum quantum number. This form gives a simple family of states characterized by a single parameter ($\kappa$), which can extrapolate between uncorrelated states $\kappa=\sqrt{N}$ and highly squeezed states approaching the $|m=0\rangle$ Fock state $\kappa\rightarrow 0$. It may be possible to identify more optimal states~\cite{andersprl2000} but this simple form is sufficient for our present purpose. We consider the limit where $N\gg1$ such that we can replace sums with integrals when calculating the moments of the angular momentum operators. This allows us to get analytical expressions for the moments of $J_z$ in equation \eqref{eq:thetanoise1}-\eqref{eq:thetanoise3}. All these terms will decrease with growing $\kappa$ since $\Delta J_{z}$ decreases for growing $\kappa$, e.g., for a coherent spin state $\kappa=\sqrt{N}$ we have $\Delta J_{z} \sim 0$. For a fixed $\kappa$ all three terms will also decrease with a growing number of measurements $n$ until a certain $n_{max}(\kappa)$ is reached. Since we have set all $\beta_{i}=1$ in our analytical calculations we find that for $n>n_{max}(\kappa)$ the noise will increase and $n_{max}(\kappa)$ is thus a minimum indicating that there is an optimal number of measurements. In our numerical simulations, however, we include the correct $\beta_{i}$'s and find that increasing $n$ above $n_{max}(\kappa)$ have no effect, i.e. there are no further noise reduction or enhancement. This is because the optimal feedback algorithm (with $\beta_{i}\neq 1$) knows not to react too strongly to measurements, which provide to little useful information. At $n_{max}(\kappa)$ the uncertainty in $\hat{J}_{z}$  prevents us from gaining information by introducing more measurements. We believe that this effect is due to the finite probability of measuring a $J_{z}$ with opposite sign than $\avg{\hat{J}_{z}}$, which spoils the measurement strategy. Note that $n_{max}(\kappa)$ grows with $\kappa$ since the width of $\hat{J}_{z}$ decreases with $\kappa$. 

Of the three therm in Eqs. \eqref{eq:thetanoise1}-\eqref{eq:thetanoise3}, the term in equation \eqref{eq:thetanoise3} is decreasing the slowest with $n$ since this contains $(1\!-\!\hat{J}_{z}/\avg{\hat{J}_{z}})$ to the lowest power.  In the following we therefore focus on this term. In order for the performance to be nearly Heisenberg limited we need the contribution from this term to be close to or smaller than then Heisenberg limit, and this put restrictions on the possible values of $\kappa$. To determine the conditions for $\kappa$ we have numerically solved the equation $\avg{(1\!-\!\hat{J}_{z}/\avg{\hat{J}_{z}})^{2n-2}\hat{J}_{z}^{2}/\avg{\hat{J}_{z}}^{2}}=1/N^{2}$ since $1/N^{2}$ is the Heisenberg limit of $\avg{\delta\Phi_{n}^{2}}$. This condition thus determines the parameters for which the noise from $\Delta J_{z}$ is comparable to the Heisenberg limit. For $N$ in the range $N=10^{3}-10^{9}$ we have found the minimum $\kappa$ for which the equation is fulfilled with $n=n_{max}(\kappa)$ i.e. assuming the optimal number of measurements. The result is that $\kappa$ needs to grow with increasing $N$ to suppress the noise in $J_z$, but the growth can be slower than $\kappa\sim\log\sqrt{N}+2$ if $n=n_{max}\sim3\log N$ measurements are used. Thus if we choose $\kappa\sim\log\sqrt{N}+2$ and $n\sim3\log N$ the noise terms in equation  \eqref{eq:thetanoise1}-\eqref{eq:thetanoise3} will decrease as $\lesssim 1/N^{2}$, which is the Heisenberg limit.

We now turn to the measurement noise, which consists of two parts. One is the accumulated back action contained in $\avg{\delta\hat{J}_{3,n}^{2}}$ (see equation \eqref{eq:estimate2}) while the other is due to the noise in the probe light (the last term in equation  \eqref{eq:estimatei}). For now we consider the accumulated back action. Using equation  \eqref{eq:noiseoperators1}-\eqref{eq:noiseoperators2} we find that this is dominated by $\sum_{i=1}^{n-1}\Omega_{i}^{2}\avg{\hat{X}_{i}^{2}}\avg{(\delta\phi_{i}^{e}\hat{J}_{x}/\avg{\hat{J}_{z}})^{2}}$ and that the dominant terms from each measurement are
\begin{eqnarray}
 i=1: &\quad&\frac{1}{2}\avg{\sin(\delta\phi_{0})^{2}}\avg{(1\!-\!\hat{J}_{z}/\avg{\hat{J}_{z}})^{2}\hat{J}_{x}^{2}/\avg{\hat{J}_{z}}^{2}}\Omega_{1}^{2}\\
 i>1: &\quad&\frac{1}{2}\avg{(\sin(\delta\phi_{0})-\delta\phi_{0})^{2}}\avg{(1\!-\!\hat{J}_{z}/\avg{\hat{J}_{z}})^{2i-2}\hat{J}_{z}^{2}\hat{J}_{x}^{2}/\avg{\hat{J}_{z}}^{4}}\Omega_{i}^{2}
\end{eqnarray}
where we have assumed that the probe light has vacuum statistics such that $\avg{\hat{P}}=\avg{\hat{X}}=\avg{\hat{X}\hat{P}}=0$ and $\avg{\hat{X}^{2}}=\avg{\hat{P}^{2}}=1/2$. Again we can get analytical expressions for $\avg{(1\!-\!\hat{J}_{z}/\avg{\hat{J}_{z}})^{2}\hat{J}_{x}^{2}/\avg{\hat{J}_{z}}^{2}}$ and $\avg{(1\!-\!\hat{J}_{z}/\avg{\hat{J}_{z}})^{2i-2}\hat{J}_{z}^{2}\hat{J}_{x}^{2}/\avg{\hat{J}_{z}}^{4}}$ by using the Gaussian approximation. Motivated by the previous numerical calculations we set $\kappa=\log\sqrt{N}+2$, $n\sim3\log N$ and by numerically evaluating the terms for $N=10^{3}\to10^{9}$ we find that for $\Omega_{i}=N^{-1+\frac{i}{n+1}}$ all terms will be $\lesssim1/N^{2}$, i.e. at the Heisenberg limit.

We now consider the part of the measurement noise that comes from the noise in the probe light. From equations  \eqref{eq:noiseoperators1}-\eqref{eq:noiseoperators2} and equation  \eqref{eq:estimatei} we find that the dominant terms are $\sum_{i=1}^{n-1}\avg{(1\!-\!\hat{J}_{z}/\avg{\hat{J}_{z}})^{2n-2i}}\avg{\hat{P}_{i}^{2}}/(\avg{\hat{J}_{z}}^{2}\Omega_{i}^{2})$. Again by numerically evaluating the terms for $N=10^{3}\to10^{9}$ we get the scaling of the terms and we find that for $\kappa=\log\sqrt{N}+2$, $n\sim3\log N$ and $\Omega_{i}=N^{-1+\frac{i}{n+1}}$ all the terms will be $\lesssim1/N^{2}$.  

So far we have found that we can make the noise from the measurements and from $\Delta J_{z}$ be $\lesssim1/N^{2}$, which is the Heisenberg limit. The limiting noise in $\avg{\Phi_{n}^{2}}$ is then the noise from $\Delta J_{y}$. From \eqref{eq:estimate2} we find that this has a contribution of  $\avg{\hat{J}_{y}^{2}}/\avg{\hat{J}_{z}}^{2}=\Delta\hat{J}_{y}^{2}/\avg{\hat{J}_{z}}^{2}$. For the states $\ket{\psi(\kappa)}$ we find that $\Delta\hat{J}_{y}^{2}/\avg{\hat{J}_{z}}^{2}\sim\kappa^{2}/N^{2}$. We thus get  $\sigma_{\gamma}\sim(2/N+\log\sqrt{N}/N)/\sqrt{\gamma T}$ for $\kappa=\log\sqrt{N}+2$, $n\sim3\log N$ and $\Omega_{i}=N^{-1+\frac{i}{n+1}}$. Note that we are in the limit of $N\gg1$ and that $\sigma_{\gamma}$ is in units of $(\gamma/(\omega^{2}\tau))^{1/2}$ (see article). For comparison the Heisenberg limit for the same Ramsey time is $\sigma_{\gamma}=(1/N)/\sqrt{\gamma T}$ in the same units. Hence our results shows that near Heisenberg limited stability can be obtained with the adaptive protocol.

We will also consider the performance of the adaptive protocol with uncorrelated atoms ($\kappa=\sqrt{N}$) . In this case as for the conventional Ramsey protocol the stability will be limited by the projection noise and will be $\sigma_{\gamma}\sim N^{-1/2}/\sqrt{\gamma T}$. Furthermore we will compare with the optimized protocol of Ref.\cite{andre2004prl} where the stability is $\sigma_{\gamma}\sim N^{-2/3}/\sqrt{\gamma T}$.  

\section{Numerical simulation}

To verify the above findings we have simulated an atomic clock with a LO subject to both white and $1/f$ noise. For both types of noise we have simulated the clock for atom numbers ranging from $100$ to $10^{6}$. For $N\leq1000$ we simulate the full quantum evolution during the measurements by bringing the input state $\ket{\psi(\kappa)}$ trough a Ramsey sequence and mixing it with a light state, which is assumed to be in vacuum. We pick the measurement outcome of $\hat{P}^{'}$ according to the corresponding probability distribution and subsequently update the state of the atoms for the next measurement etc. We denote this as 'full quantum simulation'. For $N>1000$ we approximate the probability distributions of $\hat{J}_{x,y,z}$ with Gaussian distributions with moments calculated from $\ket{\psi(\kappa)}$ in the limit of $N\gg1$ such that we can replace the sum over $m$ with an integral. We denote this as 'Gaussian simulation'.  

\begin{figure} 
\includegraphics[width=0.5\textwidth]{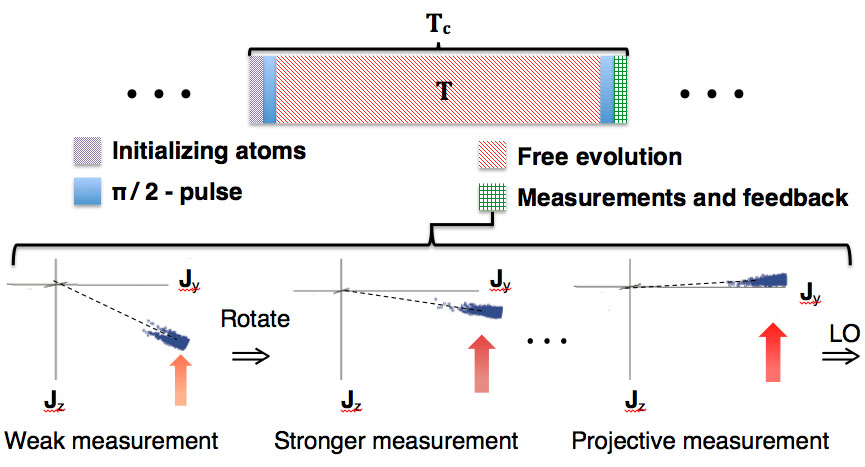} 
\caption{(Colour online) Operation of an atomic clock. A clock cycle of duration $T_{c}$ starts with initializing the atoms and ends with the measurements and feedback on the LO. We assume $T_{c}\sim T$ i.e. a negligible Dick noise \cite{dick} such that the clock is limited by the quantum noise of the atoms. The bottom part of the figure shows the adaptive protocol consisting of a series of weak measurements with intermediate feedback. The atomic spin lies in the $zy$-plane after the final $\pi/2$-pulse and the feedback seeks to rotate it to have mean spin almost along the $y$-axis before the final projective measurement and subsequent feedback on the LO.}
\label{fig:new_figure}
\end{figure}

The adaptive measurement protocol and the feedback on the LO are simulated as described in equation (1) and below in our article. The clock cycle is pictured in Fig. 2 which is reproduced as \figref{fig:new_figure}.
The total time of the clock cycle is denoted $T_{c}$. The fluctuations of the LO are however only monitored during the Ramsey time $T$ and the initialization time, the measurement time, and the time of the two $\pi/2$ pulses is thus dead time $T_{dead}$ of the clock, where we do not monitor the fluctuations of the LO. This dead time results in the Dick noise \cite{dick}, which may limit the stability of the clock. It is therefore desirable to have the ratio $T/T_{c}$ as close to unity as possible, to minimize the Dick noise, but this ratio depends on the experimental setup used to realize the clock. By technical improvements in the setups being used, e.g., by decreasing the dead time and improving the LO stability, it is in principle possible to reduce the Dick noise. Ultimately the clock will then be limited by the quantum noise of the atoms, which is denoted the standard quantum limit (SQL) for uncorrelated atoms. For current optical lattice clocks the limit of the stability is not the SQL but rather the Dick limit but a significant amount of research is put into pushing the stability towards the SQL \cite{nicholsonprl2012,lodewyck2010,takamoto2011}. Clocks based on trapped ions and atomic fountain clocks can, however, be operated with a small Dick noise and demonstrations of SQL limited clocks have been reported \cite{chouprl2010,peik2006,santarelliprl1999} (Although the fountain clock are not limited by LO decoherence and are thus less relevant for our study). Alternatively some clock based measurements can be performed which circumvent the Dick noise~\cite{takamoto2011,hinkley,lodewyck2010}. In this work we address the problem of improving a clock beyond the SQL and hence we assume negligible dead time of the clock. The adaptive measurement protocol that we propose will inevitably increase the dead time of the clock compared to a clock operated with projective measurements but this increase depends only logarithmically on $N$ since the number of weak measurement is $\sim 3\log N$. Furthermore we increase the Ramsey time of the clock by a factor of 3 for white noise in the LO. The ratio between the measurement time and the Ramsey time thus only increases with a factor proportional to $\log N$ where the proportionality constant is the ratio between the time of a weak measurement plus subsequent feedback and a projective measurement We therefore do not expect our scheme to significantly enhance the Dick effect compared to a conventional Ramsey clock operated with projective measurements. However to gain the full advantage of our protocol the Dick limit needs to be small compared to the quantum noise limited stability that we calculate. For a LO with white noise the Dick limit of the stability is $\sigma_{\gamma}\sim\sqrt{T_{dead}/T_{c}}$ in units of $[(\gamma/\omega^{2}\tau)^{1/2}]$~\cite{westergaardphd}. However  it is more realistic that the LO is subject to correlated noise such as $1/f$ noise. We have numerically simulated the Dick limited performance of our protocol in the case of $1/f$ noise in the LO and found that the stability scales as $\sim T_{dead}/T_{c}$. For $N=10^{6}$ atoms we find a quantum noise limited stability of $\sigma_{\gamma}\sim10^{-5}$ in units of $[(\gamma/\omega^{2}\tau)^{1/2}]$ from our protocol whereas optimal squeezing for the standard Ramsey spectroscopy \cite{andre2004prl} gives $\sigma_{\gamma}\sim3\cdot10^{-4}$ . From our numerical simulation of the Dick limited stability we find that for this number of atoms the standard protocol with squeezed states is limited by atomic noise for $T_{dead}\lesssim3\cdot10^{-5}T_{c}$, and hence the protocol described here can be used to improve the stability in this regime. To gain the full advantage of the protocol we would need $T_{dead}\lesssim10^{-6}T_{c}$. For lower number of atoms the current protocol would be advantageous at higher values of $T_{dead}$ e.g. at $N=100$ the standard protocol is limited by atomic noise for $T_{dead}\lesssim2\cdot10^{-2}T_{c}$ and we gain the full advantage of our protocol for $T_{dead}\lesssim10^{-2}T_{c}$.      

\begin{figure} [H]
\centering
\subfloat {\label{fig:whitenoise}\includegraphics[width=0.4\textwidth]{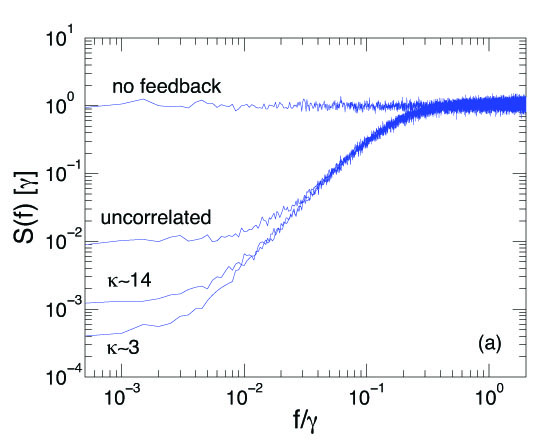}} 
\subfloat{\label{fig:pinknoise}\includegraphics[width=0.4\textwidth]{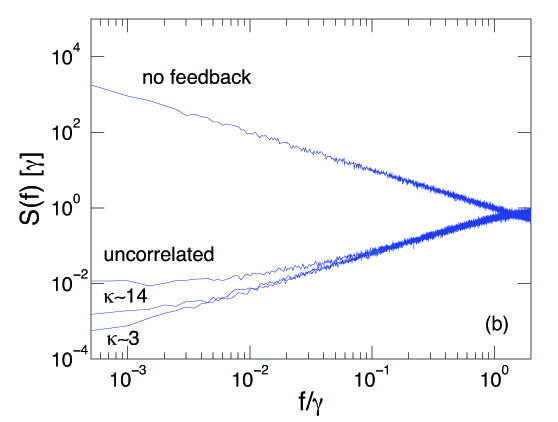}}
\caption{(Colour online) Noise spectrum of the slaved LO for (a) white noise and (b) $1/f$-noise. The noise spectrum is defined as $S(f)\delta(f+f')=\avg{\delta\omega(f)\delta\omega(f')}$ where $\delta\omega(f)$ is the Fourier transform of the frequency fluctuations $\delta\omega(t)$ of the LO. The plots show how the feedback effectively locks the LO to the atoms so that the noise in the LO becomes limited by the atomic noise for low frequencies.  Furthermore it is seen how squeezing lowers the atomic noise and hence the noise of the locked LO. The plots were made for $N=1000$ and $\gamma T=0.1$. Counting from above the curves show the noise spectrum for a unlocked LO, the conventional protocol with uncorrelated atoms, the conventional protocol with the optimal squeezing of $\kappa\sim14$ and the adaptive protocol with optimal squeezing of $\kappa\sim3$. Note that the optimal performance of the adaptive protocol is reached at higher $\gamma T$ and further improvement is thus possible.}
\label{fig:noisespec}
\end{figure}

Assuming negligible Dick noise the feedback effectively locks the LO to the atoms, and lowers the noise level of the LO as shown in \figref{fig:noisespec} where the noise spectrum $S(f)$ of the LO is plotted against the frequency $f$. The noise spectrum is here defined as $S(f)\delta(f+f')=\avg{\delta\omega(f)\delta\omega(f')}$ where $\delta\omega(f)$ is the Fourier transform of the frequency fluctuations $\delta\omega(t)$ of the LO. For a free running LO with white noise we use $S(f)=\gamma$ while for 1/f noise we use $S(f)=\gamma^{2}/f$ where $\gamma$ is a parameter characterizing the fluctuations of the LO. The figure show that for high frequencies the locked LO has  the noise of the free running oscillator but for low frequencies the LO is locked to the atoms and is limited by the atomic noise.  
Furthermore \figref{fig:noisespec} shows how squeezing improves the stability of the clock by lowering the noise level of the locked LO more than for uncorrelated atoms. While the conventional Ramsey scheme works ideally  for $\kappa\sim 14$, the adaptive protocol allows for $\kappa \sim 3$ at the atom number $N=1000$ used in the figure, and thus leads to an improved stability. 

For a fixed Ramsey time the feedback strength ($\alpha$) determines how long time the clock has to run before the LO is locked to the atoms. Since we simulate a clock running for a long but finite time there will be some remaining information from the last measurement results, which have not been fully exploited by the feedback loop. In our simulations we therefore do an additional phase correction to the LO after the final measurement. In principle the influence of the last few measurements could also have been reduced by running the simulation for a longer time, but by doing the correction we reduce the required simulation time. To find the required phase correction and obtain an expression for the stability of the clock we study the phase of the locked LO. At time $t_{k}=kT$ the phase of the LO is
\begin{equation}
\delta\phi(t_{k})=\int_{t_{k-1}}^{t_{k}}{\left( \delta\omega_{0}(t)+\sum_{i=1}^{k-1}\Delta\omega_{i}\right)} \text{d}t,
\end{equation}
where $\delta\omega_{0}(t)$ is the frequency fluctuations of the unlocked LO and $\Delta\omega_{i}$ is the frequency corrections applied at time $t_{i}$. Using that $\Delta\omega_{i}=-\alpha\delta\phi_{e}(t_{i})/T$ where $\delta\phi_{e}(t_{i})$ is the estimated phase of the LO at time $t_{i}$ we can write
\begin{equation} \label{eq:phase}
\delta\phi(t_{k})=\delta\phi_{0}(t_{k})-\alpha\sum_{i=1}^{k+1}\delta\phi_{e}(t_{i}),
\end{equation}
where $\delta\phi_{0}(t_{k})=\int_{t_{k-1}}^{t_{k}} \delta\omega_{0}(t) \text{d}t$. The mean frequency offset of the LO after running for a period $\tau=lT$ ($l \gg 1$) is
\begin{equation} \label{eq:offset1}
\delta\bar{\omega}(\tau)=\frac{1}{\tau}\left(\sum_{i=1}^{l}\delta\phi(t_{i})-\phi_{\text{final correct}}\right),
\end{equation}  
where $\phi_{\text{final correct}}$ is the phase correction that we apply after the final measurement. Combining equation \eqref{eq:phase} and \eqref{eq:offset1} we find that $\phi_{\text{final correct}}=\sum_{i=1}^{l}\left((1-\alpha)^{l-i}\delta\phi_{e}(t_{i})+\sum_{j=1}^{i-1}\alpha(1-\alpha)^{l-i}\delta\phi_{e}(t_{j})\right)$ will give the ideal performance. For this choice of $\phi_{\text{final correct}}$ the mean frequency offset becomes 
\begin{equation}
\delta\bar{\omega}(\tau)=\frac{1}{l}\sum_{i=1}^{l}\frac{\delta\phi(t_{i})-\delta\phi_{e}(t_{i})}{T},
\end{equation}
i.e. the error is determined by the sum of the phase estimation errors. We use this expression to determine the stability of the clock, which is given by $\sigma_{\gamma}(\tau)=\avg{(\delta\bar{\omega}(\tau)/\omega)^{2}}^{1/2}$. Note that while the sum is over different time intervals we cannot in general determine the stability by looking at different intervals independently since the phases are correlated for finite $\alpha$ or for correlated noise in the free running LO e.g. $1/f$ noise. In our analytical calculations, however, we assume white noise and $\alpha\rightarrow 0$ so that we can ignore the correlations and consider each Ramsey sequence independently.

\begin{figure} [H]
\centering
\subfloat {\label{fig:ramseytime_white}\includegraphics[width=0.4\textwidth]{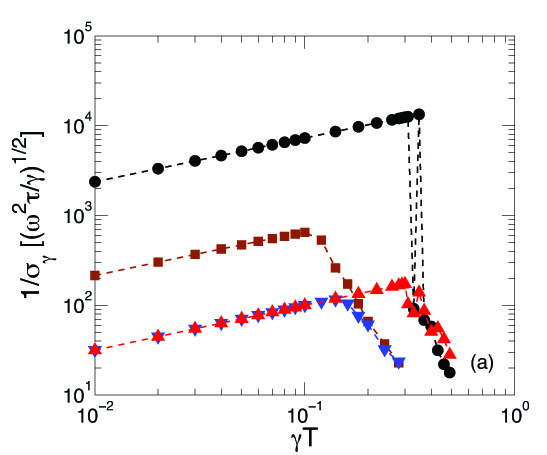}} 
\subfloat{\label{fig:ramseytime_pink}\includegraphics[width=0.4\textwidth]{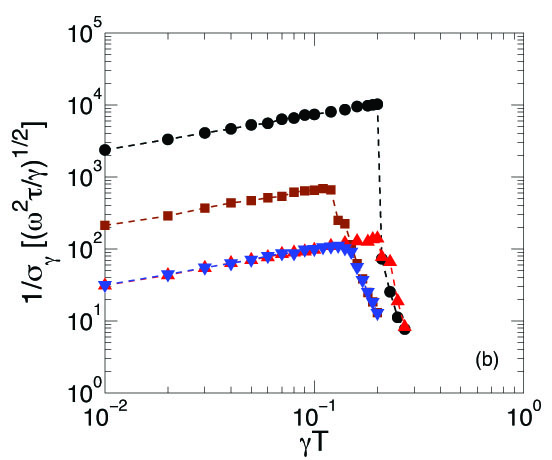}}
\caption{Stability as a function of the Ramsey time ($\gamma T$) for (a) white noise and (b) $1/f$ noise in the LO. The plots were made with $N=10^{5}$. $ \blacksquare,\blacktriangledown$ are the best non-linear protocol of Ref.~\cite{andre2004prl} while $\bullet,\blacktriangle$ are the adaptive protocol discussed in this article. The adaptive protocol allows for $\gamma T \sim 0.3$ and $0.2$ for white and $1/f$ - noise respectively while the conventional protocol of Ref.~\cite{andre2004prl} only allows for $\gamma T \sim 0.1$ for both white and $1/f$ noise. $\blacktriangledown,\blacktriangle$ correspond to uncorrelated atoms while $\bullet,\blacksquare$ are the ideal choices of squeezing in the adaptive and conventional protocols respectively. The probabilistic nature of the errors that limits $\gamma T$ is visible in (a) where the stability for the adaptive protocol jumps back and forth for  $\gamma T > 0.3$ before it is definitely diminished. }
\label{fig:noise}
\end{figure}

For the model investigated here, the stability increases with the Ramsey time $T$, but $T$ is limited by two types of errors. For experiments or simulations running with a fixed Ramsey time there will always be a finite probability that the feedback loop jumps to a state with a phase difference of $2\pi$ (so called {\it fringe hops}~\cite{rosenband2012arxiv}) or that a phase jump  that is large enough to spoil the measurement strategy occurs. For the adaptive scheme this happens for phase jumps  $\gtrsim \pi$ while it happens for phase jumps $\gtrsim \pi/2$ for the conventional protocol. The reason for this is that the adaptive protocol is able to distinguish whether the phase lies in the intervals  $[0;\pi/2]$ or $[\pi/2;\pi]$ since they will lead to different responses when we rotate the state during the feedback. On the contrary  the conventional protocol only has a single projective measurement and cannot distinguish in which of the two intervals the phase lies. In our simulations we see the phase jumps as an abrupt break down as we increase the width of the distribution of the acquired phase, i.e. as we increase $T$ since the variance is $\sigma^{2}=\gamma T$ for white noise. This break down is clearly visible in Fig.~1b (reproduced as part (a) of \figref{fig:noise} where we also show the similar plot for $1/f$ noise (b)). Below we will investigate this breakdown in more detail.    

For a simulation running for a time $\tau=lT$, we have $l$ samples of the accumulated phase of the LO during the Ramsey time $T$. Assuming that these samples have an independent Gaussian probability distribution with zero mean (this will be the case for white noise and is approximately true for $1/f$ noise when the LO is locked to the atoms), the probability of all of these phases to be less than a critical value ($a$) is
\begin{equation}
P(\le a)=\left(1-\text{erfc}\left(\frac{a}{\sqrt{2}\sigma}\right)\right)^{l}
\end{equation}
where the variance of the distribution $\sigma$ depends on $\gamma T$. For large $l$ and as a function of $\sigma$ this probability will drop abruptly from $\sim 1$ to $\sim 0$ around a certain $\sigma=\sigma_{max}$. Defining $\sigma_{max}$ to be the position where $P(\le a)=1/2$ we find that
\begin{equation} \label{eq:sigmamax}
1/2=\left(1-\text{erfc}\left(\frac{a}{\sqrt{2}\sigma_{max}}\right)\right)^{l} .
\end{equation}  
Solving this equation gives
\begin{equation} \label{eq:last}
\sigma_{max}\approx\frac{a}{\sqrt{\ln(2/\pi)+2\ln(l)-2\ln(\ln{2})-\ln(\ln(2/\pi)+2\ln(l)-2\ln(\ln{2}))}},
\end{equation} 
where we have expanded to first order in $z=(1-2^{-1/l})\sim\ln(2)/l$. It is seen that the breakdown ($\sigma_{max}$) has a weak (logarithmic) dependence on $l$. Solving equation \eqref{eq:sigmamax} with the lhs. being equal to 0.95 and 0.05 we find that  $P(\le a)$ drops from 0.95 to 0.05 within a window of $\sim 2\sigma_{max}/\ln(l)$. Hence for large $l$ the errors will appear very abruptly in the simulations. 

Ideally we should include correction strategies for the errors due to  large phase jumps in our simulations (e.g. running with different Ramsey times would correct for fringe hops), but for simplicity we ignore this. This means that our simulations have a weak dependence on the number of steps we simulate, but since \eqref{eq:last} shows that it is only a logarithmic correction we do not expect this to change our results significantly. Instead we find  the upper limit of $\gamma T$ from the simulations plotted in \figref{fig:noise}, where $l=10^6$ (for $1/f$ noise we average over 100 independent runs with $l=10^{4}$). 

When minimizing the stability for the adaptive protocol we numerically minimize in both the degree of squeezing, the number of weak measurements and the strengths of the measurements (we also minimize in the degree of squeezing for the conventional protocol of Ref. \cite{andre2004prl} for comparison with the adaptive protocol).  For white noise we can use the expression for $\sigma_{\gamma}$ in the limit of $\alpha \ll 1$. For $1/f$ noise we include the feedback on the LO as described in equation  (2) and below in the article. Note that $0<\alpha<1$ in order for the feedback to stabilize the LO \cite{andrephd}. We have used $\alpha=0.1$ in our simulations and do not expect our results to change significantly for a different choice of $\alpha$. As previously mentioned $\alpha$ determines how long time it takes for the feedback to lock the LO to the atoms (the LO is locked after a time $\sim T/\alpha$). As long as the long term stability is considered at a time $\tau \gg T/\alpha$ the LO is effectively locked to the atoms and the stability does not depend on $\alpha$ (as mentioned above the final phase correction also correct for the influence of the last measurements). To support this we have repeated the simulations of \figref{fig:ramseytime_pink} with $\alpha=0.5,0.8,$ and $0.9$. The simulations show basically the same limits to $\gamma T$ as seen in  \figref{fig:ramseytime_pink} and more or less identical results for the long term stability for $\gamma T$ below these limits, which supports the above analysis. For $\gamma T$ exceeding the maximal limit for the adaptive protocol the phase jump errors diminish the stability even more than shown in \figref{fig:ramseytime_pink} for larger $\alpha$, i.e. stronger feedback. This is because the probability of a phase jump error to result in a fringe hop is greater for a stronger feedback. For the conventional protocol we do not see this effect right above the maximal limit of $\gamma T$ because the probability of a phase jump error to result in a fringe hop is small. However we expect to see the same effect if we increased $\gamma T$ well above the maximal limit of the conventional protocol.�

\end{document}